\documentclass[sigconf, screen]{acmart}

\renewcommand\footnotetextcopyrightpermission[1]{}
\author{Jia Li}
\affiliation{%
  \institution{The Chinese University of Hong Kong}
  \country{}
  }
\email{linsayli@link.cuhk.edu.hk}

\author{Jiacheng Shen}
\affiliation{%
  \institution{Duke Kunshan University}
  \country{}
  }
\email{jc.shen@dukekunshan.edu.cn}

\author{Yuxin Su}
\affiliation{%
  \institution{Sun Yat-sen University}
  \country{}
  }
\email{suyx35@mail.sysu.edu.cn}

\author{Michael R. Lyu}
\affiliation{%
  \institution{The Chinese University of Hong Kong}
  \country{}
  }
\email{lyu@cse.cuhk.edu.hk}
\AtBeginDocument{%
  \providecommand\BibTeX{{%
    \normalfont B\kern-0.5em{\scshape i\kern-0.25em b}\kern-0.8em\TeX}}}

\usepackage{multirow}
\usepackage{tikz}
\usepackage{amsmath}
\usepackage{xspace}
\usepackage{enumitem}
\usepackage{booktabs}
\usepackage{bbding}
\usepackage{subfig}
\usepackage{stfloats}
\usepackage{indentfirst}
\usepackage{makecell}
\usepackage[font=small]{caption}
\newcommand{\ie}{{\em i.e.},\xspace}
\newcommand{\eg}{{\em e.g.},\xspace}

\usepackage{ listings}
\usepackage{xcolor}
\definecolor{mygreen}{rgb}{0,0.6,0}
\definecolor{mygray}{rgb}{0.5,0.5,0.5}
\definecolor{mymauve}{rgb}{0.58,0,0.82}
\begin{document}

\title{ColorGo: Directed Concolic Execution}




\begin{abstract}
    Directed fuzzing is a critical technique in cybersecurity, targeting specific sections of a program. This approach is essential in various security-related domains such as crash reproduction, patch testing, and vulnerability detection. Despite its importance, current directed fuzzing methods exhibit a trade-off between efficiency and effectiveness. For instance, directed grey-box fuzzing, while efficient in generating fuzzing inputs, lacks sufficient precision. The low precision causes time wasted on executing code that cannot help reach the target site. Conversely, interpreter- or observer-based directed symbolic execution can produce high-quality inputs while incurring non-negligible runtime overhead. These limitations undermine the feasibility of directed fuzzers in real-world scenarios. 

    To kill the birds of efficiency and effectiveness with one stone, in this paper, we involve compilation-based concolic execution into directed fuzzing and present ColorGo, achieving high scalability while preserving the high precision from symbolic execution.
    ColorGo is a new directed whitebox fuzzer that concretely executes the instrumented program with constraint-solving capability on generated input. It guides the exploration by \textit{incremental coloration} including static reachability analysis and dynamic feasibility analysis. 
    We evaluated ColorGo on diverse real-world programs and demonstrated that ColorGo outperforms AFLGo by up to \textbf{100×} in reaching target sites and reproducing target crashes. 
\end{abstract}

\maketitle

\section{Introduction}\label{sec:introduction}
Directed fuzzing is an approach that aims to reach a target site of a program under test, \eg a target line of code, by iteratively generating inputs named seeds.
Due to its directed nature, it plays a vital role in various software testing and debugging tasks, \eg patch testing~\cite{marinescu2013katch, godefroid2005dart}, crash reproduction~\cite{jin2012bugredux,du2022windranger,kim2019poster}, and vulnerability detection~\cite{wen2020memlock, wang2020typestate, petsios2017slowfuzz}.
For direct fuzzers, the time it takes to reach a target site is a key performance metric. 
However, existing approaches, \ie graybox fuzzers~\cite{AFLGo,du2022windranger,chen2018hawkeye,huang2022beacon,luo2023selectfuzz}, and whitebox ones~\cite{jin2012bugredux,marinescu2013katch}, suffer from poor efficiency due to their slow or imprecise seed generation process.

The state-of-the-art directed fuzzing is directed graybox fuzzing, which achieves high seed-generating speed~\cite{AFLGo}. 
However, graybox approaches suffer from poor performance due to the low precision of the generated seeds since new seeds are generated by randomly mutating existing ones.
Many useless seeds are generated in the random process, wasting execution time.
For example, as shown in Figure~\ref{fig:long-condition}, if we want to satisfy the condition $input=123456790$ in a graybox fuzzer, we need to mutate at most $2^{32}$ times, any intermediate outcome is considered useless.

Another approach is directed whitebox fuzzing.
Whitebox approaches are based on symbolic execution.
They can leverage the internal structure of the program and generate precise solutions by solving constraints.
However, existing whitebox approaches rely on interpretation-based symbolic engines, which incur high runtime overhead~\cite{cadar2008klee}.

\

\begin{lstlisting}[caption=Mutation Example,label=fig:long-condition, frame=shadowbox]
int input;
if (input==123456789) {
    assert(0); /*error*/
}
\end{lstlisting}


In this paper, we propose using compilation-based concolic execution to achieve directed fuzzing, \ie Directed Concolic Execution (DCE). This approach addresses the low-efficiency issue of the graybox approach and the low-precision issue of the whitebox approach.
Compared to graybox approaches, DCE can generate precise seeds by solving constraints in the path conditions.
Compared to whitebox approaches, DCE reduces runtime overhead by moving the interpretation overhead to compilation time through instrumentation.

However, two challenges have to be addressed to achieve high-performance and practical directed concolic execution.

\textbf{\textit{The lack of global information.}}
An efficient DCE needs the assistance of global information, \ie the internal structure of the program and the global runtime information, to restrict the search space and implement effective search strategy~\cite{luo2023selectfuzz,huang2022beacon}.
However, concolic execution is concretely executed over an input, which only maintains the current execution status.
As a result, a naive DCE cannot effectively direct the execution of the program, causing high search overhead or failed searches.



\textbf{\textit{The intrinsic drawbacks of symbolic execution.}}
While using symbolic execution can improve the precision of input generation, we still need to address the inherent challenges of symbolic execution. 
For instance, the loop statement introduces a state explosion problem due to its circular execution flow. Interprocedural analysis necessitates additional effort to achieve data-sensitive and control-flow-sensitive characteristics. Furthermore, the indirect call requires an adapted points-to-analysis specifically designed for the LLVM framework.
These problems can cause DCEs to be stuck or lost in the middle of a program, and thus lead to failed path-finding.

We present ColorGo, a whitebox directed fuzzer that overcomes these challenges. 
First, to gather global information, we utilize the code structure information provided by the compiler as static information. The compiler obtains the internal structure of the program during the process of analyzing and translating the source code. We use this static information to limit the search scope in terms of inter-procedural control-flow graph (iCFG) reachability, which we refer to as static coloration. This static process is completed during compilation, eliminating any runtime overhead.
Next, to supplement the global information with the runtime information from concolic execution, we perform incremental coloration at runtime, focusing on the feasibility of path constraints.
Our goal is to reduce the exploration space and avoid unnecessary searches. Once the coloration is completed, we employ early stopping and deviation basic block identification as part of our proposed efficient search strategy.
Finally, to address the inherent limitations of symbolic execution, we specifically target them based on the characteristics of concolic execution, \ie target line feedback, partial function model, and reverse edge stopping.

To evaluate the effectiveness of our design, we implement ColorGo on top of the LLVM framework. 
We compare it with state-of-the-art directed fuzzers and evaluate it on three types of real-world programs.
Our experiments show that ColorGo achieves 50×, 100× speedup for reaching target sites and reproducing vulnerabilities. 
Besides, we conduct an ablation study and shows the effectiveness of individual components in our design.

In summary, we make the following contributions in this paper:
\begin{enumerate}
    \item We propose a directed whitebox fuzzer that combines lightweight program analysis and compilation-based concolic execution to efficiently generate input for reaching specific code regions.
    \item We implement a practical system called ColorGo on top of the LLVM framework, which addresses the inherent limitations by combining the characteristics of concolic execution, achieving both high precision and scalability.
    \item We conduct experiments on real-world programs (jasper, lame, binutils), demonstrating significant performance improvements compared to the state-of-the-art directed graybox fuzzers.
\end{enumerate}

In the rest of the paper, we first elaborate on our idea of compilation-based Directed Concolic Execution (Section~\ref{sec:design}). We then present ColorGo in detail (Section~\ref{sec:implementation}) and compare its performance with state-of-the-art implementation (Section~\ref{sec:evaluation}), showing that it is orders of magnitude faster than the benchmark in testing real-world software.
\section{Background and Motivation}\label{sec:background}
In this section, we introduce two commonly used techniques for directed fuzzing, \ie directed graybox fuzzing and directed symbolic execution, and discuss their limitations. 
We then introduce our motivations to use concolic execution to overcome these limitations.
Finally, we discuss the new challenges of achieving directed concolic execution.

\subsection{Backgroud}

\indent \textbf{Directed Graybox Fuzzing.} 
Directed grabox fuzzing is the most widely adopted approach in the literature of directed fuzzing.
The fuzzing process of graybox approaches can be divided into two phases, \ie exploration and exploitation. 
During the exploration phase, a graybox fuzzer covers as many program paths as possible by iteratively mutating seeds that trigger new paths.
After a user-specific time period, the fuzzer enters the exploitation phase to focus on specific code areas. 
Specifically, graybox fuzzers use lightweight instrumentation methods to calculate the quality of seeds, \eg distance~\cite{AFLGo,du2022windranger} and similarity~\cite{chen2018hawkeye,liang2019sequence}. 
Intuitively, if a seed executes on a path that is closer to the target site, then seeds generated from it are also more likely to be close to the target.
Therefore, existing graybox fuzzers give high-quality seeds higher priorities for mutation, and generate more inputs from them. 

However, directed graybox fuzzers suffer from poor performance because of the imprecise seed generation. 
Seeds with high priorities are selected for random mutation since they will possibly generate new seeds that can satisfy the desired path conditions. 
Unfortunately, a lot of seeds that do not help promote directed fuzzing are generated due to the inaccurate priority and the randomness of seed mutation. 
These seeds lead to irrelevant execution, which is time-consuming and reduces the fuzzing performance.

\textbf{Directed Symbolic Execution.}
Different from the randomized directed graybox fuzzing, directed symbolic execution (DSE) precisely generates inputs.
It casts directed fuzzing to a step-by-step constraint-solving process. 
By thoroughly analyzing the program and extracting structural information, DSE can determine which constraint to solve and generate input that is closer to the target by solving this constraint.

However, existing approaches~\cite{marinescu2013katch,jin2012bugredux} rely on interpretation-based symbolic execution engines, \eg KLEE~\cite{cadar2008klee}, which suffer from high state management overhead.
Specifically, these symbolic execution engines are virtual machines over LLVM bitcode.
They iteratively fetch each instruction, execute the instruction symbolically, and update the symbol states in the memory model.
During the process, they fork the execution trace on each branch condition, generating huge execution states.
The heavy virtual machine and state management mechanism incur heavy runtime overhead on both computation and memory. Meanwhile, The forking system needs to manage massive information for each execution state, which causes the problem of state explosion.

\subsection{Motivation}
\subsubsection{Directed Concolic Execution}
\

\noindent To address the aforementioned problems, and take both precision and efficiency into account, our method adopts compilation-based concolic execution. 
Concolic Execution is similar to dynamic symbolic execution, as it executes concrete execution over one input and conducts analysis only over the explored path. 
To introduce symbolic characteristics, concolic execution treats variables related to input as symbolic variables and maintains corresponding symbolic expressions.
To achieve efficient symbolic execution, concolic execution does not employ a global manager that records a vast space of states. 
Instead, concolic execution implicitly maintains concrete states through the native CPU. After execution, all statuses of an execution trace are recorded as the new input. 
This stateless implementation reduces computational complexity and memory space usage, resulting in unprecedented efficiency close to native execution and fundamentally eliminating the state explosion problem. 
However, the stateless nature also introduces the problem of lack of global information, which is used to guide the fuzzing process.

There are primarily two methods to instrument the program in concolic execution, either at compilation time, \eg through an LLVM pass, like SYMCC, or at execution time using dynamic binary translators (DBT), like SymQEMU and QSYM. DBT does runtime code manipulation when a program executes, works like an observer between application and operating system, and performs JIT translation, which incurs a non-negligible overhead at runtime. For instance, as SymFusion\cite{coppa2022symfusion} mentioned, SymQEMU could be 6.5× slower than SymCC on a simple code snippet.
Our work is based on compilation-based concolic execution for its runtime performance advantage. 
We instrument the program under test at the level of the compiler's intermediate representation, which allows us to bypass the complex semantics of the source code. Consequently, our work is compatible with all source languages that can be compiled to the intermediate representation.
As Katch proposed, symbolically interpreting the program is several orders of magnitude slower than native execution, while the instrumented programs have a comparable execution time to their native counterparts. 
To achieve high precision, compilation-based concolic execution only needs one concrete execution, at the cost of one irrelevant execution, but reduces massive irrelevant explorations caused by imprecise inputs.
Besides, the code analysis results provided by the compiler offer us a convenient way to access global information.

Directed Concolic Execution, a directed fuzzing tool built upon compilation-based concolic execution, which addresses both the issues of low precision and low scalability, is a key component of our approach. 
However, the design is not as straightforward as the concept. Complex program semantics present challenges in tracking symbolic expression, and there are some inherent limitations of symbolic execution, such as handling indirect calls, interprocedural analysis, and loop unrolling.

\subsubsection{Problem statement}\label{sec:problem}
\

\noindent In this section, we provide some examples of limitations of concolic execution to expand the challenge we mention in Section~\ref{sec:introduction}. Before this, We need to illustrate the definition of relevant code. Given some targets, we identify the relevant code according to the reachability on the inter-procedure control ﬂow graph and the feasibility of the path constraint. The search scope is all relevant code.

\begin{figure}[t]
    \vspace{-0.15in}
    \centering
    \subfloat[C source Code.]{
        \includegraphics[width=0.6\columnwidth]{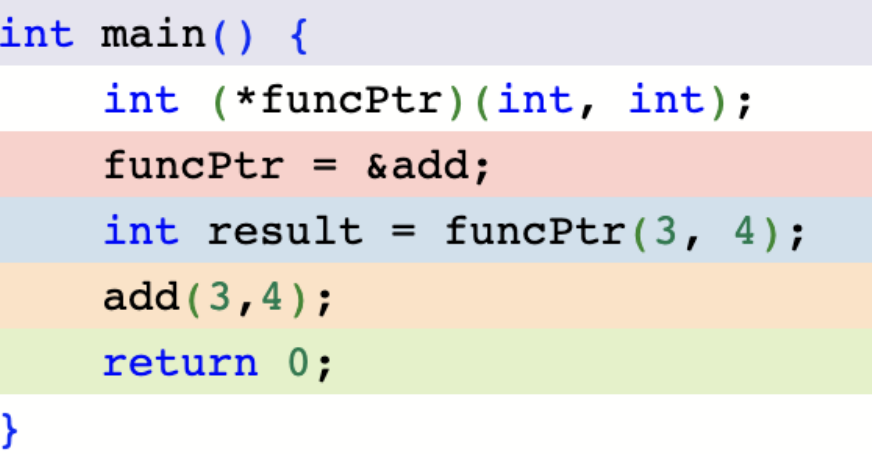}
        \label{fig:indirect-call-code}
    }
    \hspace{1mm}
    \subfloat[LLVM IR.]{
    \includegraphics[width=0.9\columnwidth]{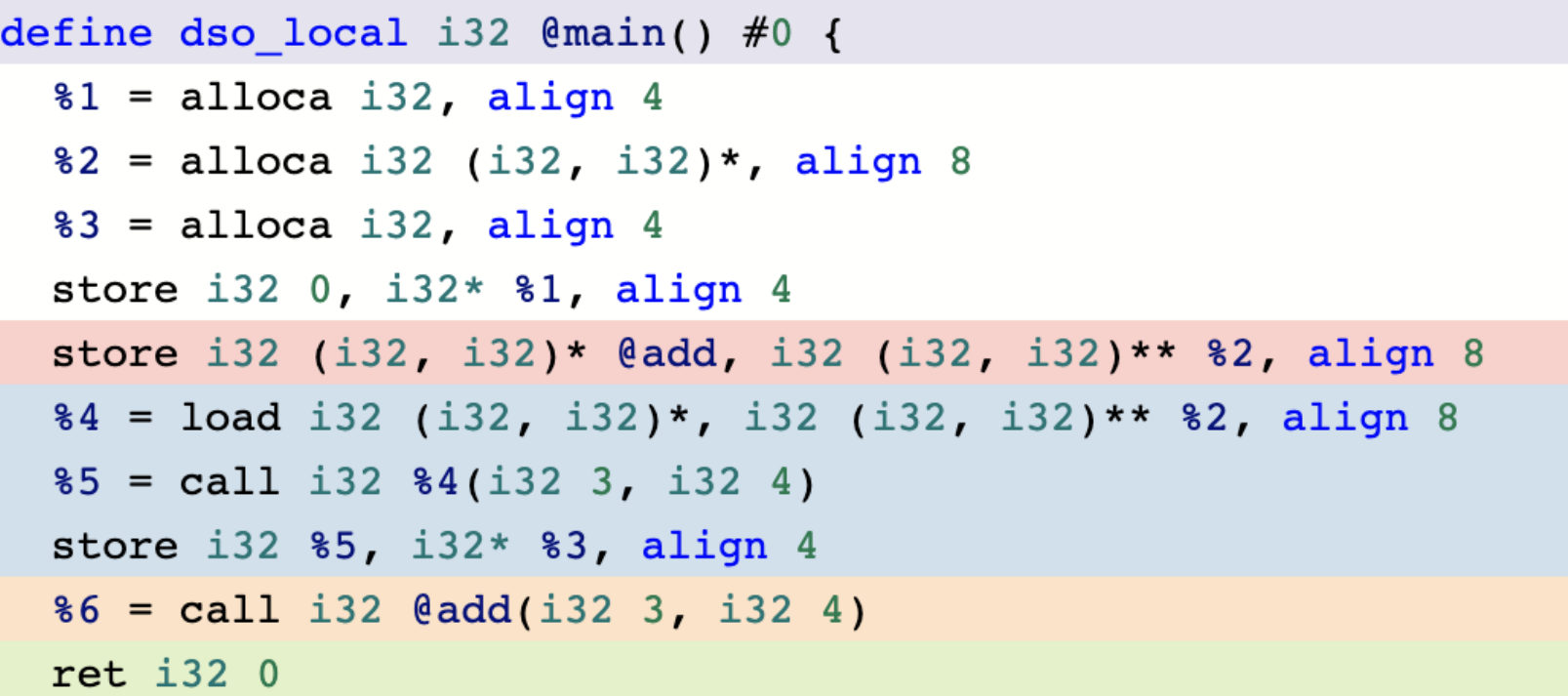}
        \label{fig:indirect-call-ir}
    }
    \caption{A indirect call example.}\label{fig:ic}
    \vspace{-0.1in}
\end{figure}

\indent \textbf{Indirect Call.}
A call graph is utilized to determine the reachability of each basic block and target site, and it is combined with the control flow graph to construct an inter-procedure control flow. Constructing an Interprocedural Control Flow Graph (iCFG) poses a significant challenge, mainly because it's challenging to infer the targets of indirect control transfer instructions, particularly indirect calls. Indirect calls are calls through register or memory operands. While modern static analysis tools, such as the SVF~\cite{sui2016svf}, can support indirect call target inference when constructing control flow graphs using points-to analysis, the LLVM compiler cannot natively support this feature.

Take Figure~\ref{fig:ic} as an example, as Figure~\ref{fig:indirect-call-code} shows, the $add$ function is called indirectly first through a function pointer $funcPtr$, and the corresponding IR shows the pointer variable $funcPtr$ is stored in memory. When we process the instruction call in Line 8 in Figure~\ref{fig:indirect-call-ir}, we can not directly resolve the indirect instruction operand to get the actual target $add$. In contrast, for the instruction call in Line 10 in Figure~\ref{fig:indirect-call-ir} which conducts a direct all, we can determine that the target function is $add$ directly without additional analysis.

That is, although we construct an iCFG involving indirect call through additional points-to-analysis, and find a path from function $main$ to $add$ (function $main$ indirectly calls function $add$) in the iCFG, we can not color irrelevant basic block correctly at compilation time, because we can not location the call instruction in $main$ whose operand is the function $add$.

\begin{figure}[t]
    \vspace{-0.15in}
    \centering
    \subfloat[Example of data-sensitive analysis.]{
        \includegraphics[width=0.45\columnwidth]{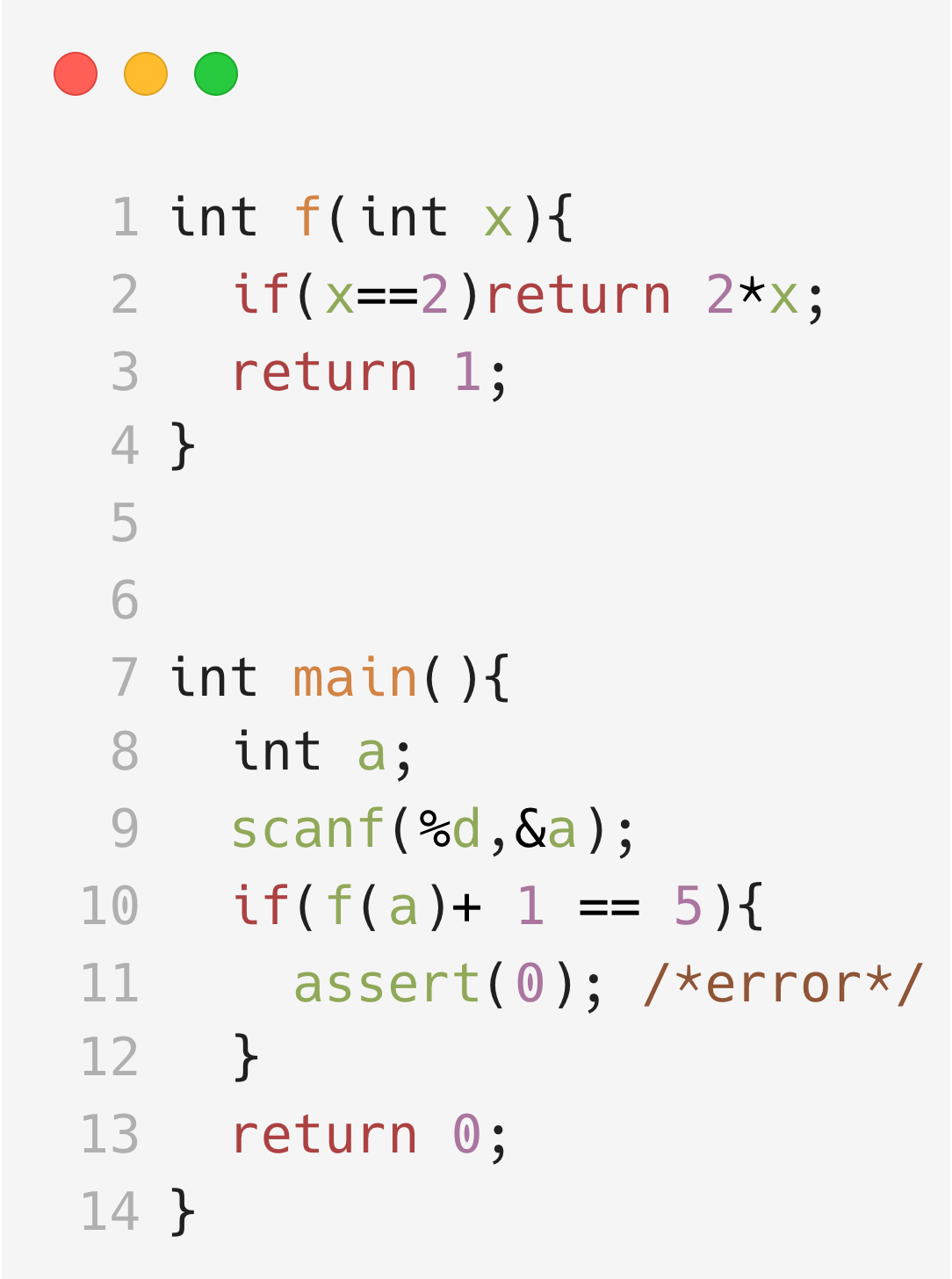}
        \label{fig:fc-data}
    }
    \hspace{1mm}
    \subfloat[Example of control-sensitive analysis.]{
        \includegraphics[width=0.45\columnwidth]{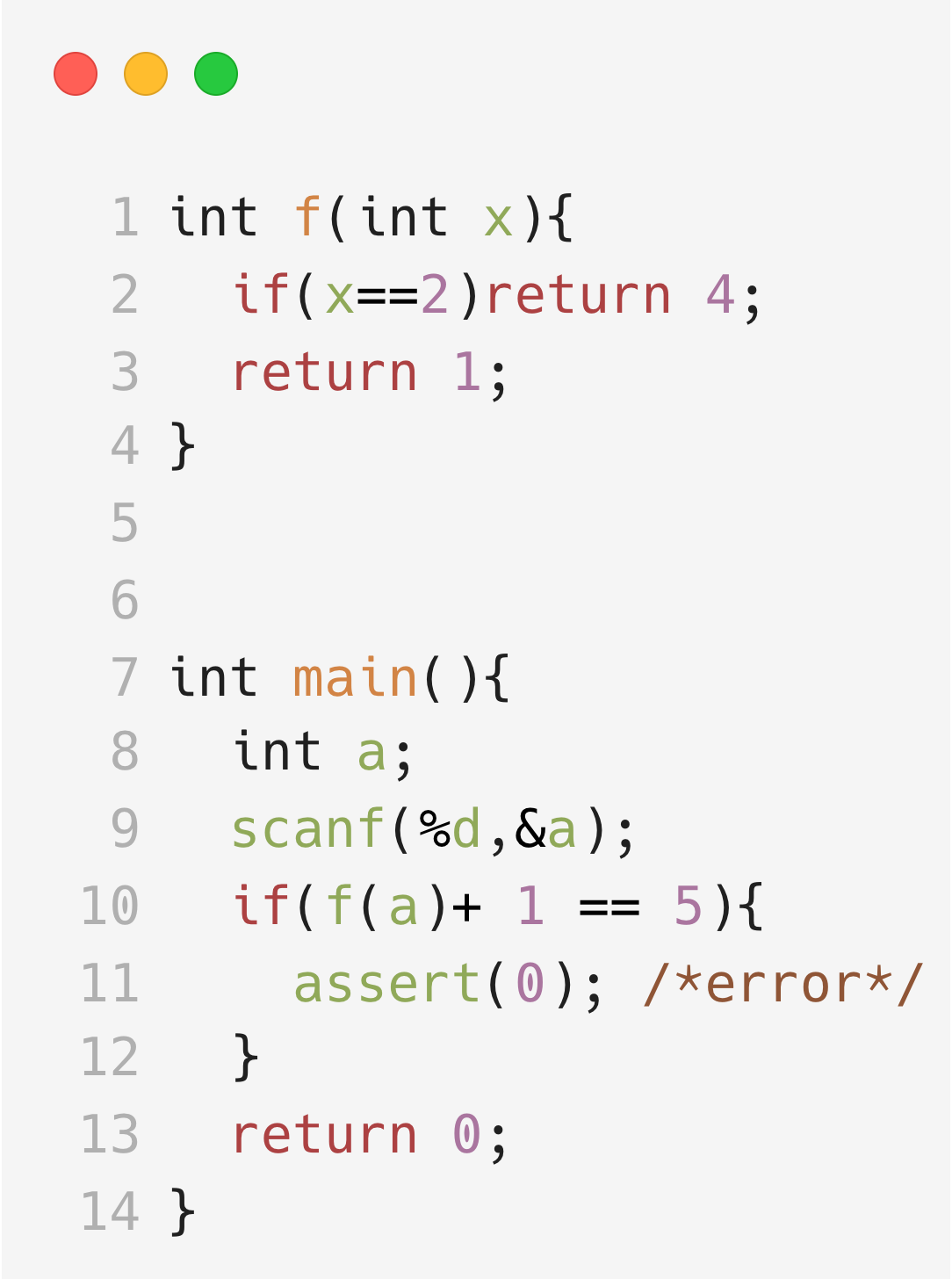}
        \label{fig:fc-control}
    }
    \caption{A function call example.}\label{fig:fc}
    \vspace{-0.1in}
\end{figure}

\indent \textbf{Interprocedural Analysis.}
Symbolic execution can be categorized into two types of analysis: intraprocedural analysis and interprocedural analysis. Intraprocedural analysis only considers statements within a procedure, whereas interprocedural analysis requires the inclusion of procedure calls to conduct whole-program analysis. Interprocedural analysis can be implemented in two ways, data-flow sensitive inter-procedural analysis and control-flow sensitive inter-procedural analysis.

Let's consider Figure~\ref{fig:fc-data} as an example. In order to reach target line 11, we need to solve the constraint in Line 10, which issues a sub-procedure call. However, the intraprocedural analyzer is unable to analyze the operations performed by the function $f()$. As a result, it replaces the symbolic value of $f(a)$ with its concrete value. Therefore, the path constraint that we formulate becomes $constant + 1 == 5$. Without any symbolic variables, no test input is generated for this equation during execution. As a result, it fails to reach the branch in Line 10, and so does the directed fuzzing. When we conduct data-flow sensitive inter-procedural analysis, the symbolic variable can pass through the statements. Because of the branch statement in Line 2, the return value depends on the concrete value of $x$. For example, if the condition $x!=2$ is true, the return value is a concrete value of 1, otherwise, the return value is a symbolic value of $2x$; Unfortunately, we only need to solve the constraint $f(a)+1!=5$ when it is not satisfied, that is, the return value of $f$ is not equal to 4, thus the value of x is not equal to 2, then we run to Line 3, return a concrete value of 1. We do not have an opportunity to solve the constraint $2*a+1==5$ and generate the wanted input $a=2$. However, if Line 2 did not have a branch and $f$ directly returned $2*x$, the situation would be simpler and solvable. Consequently, data-flow sensitive inter-procedural analysis would handle this scenario well. In this case, we can observe how control flow affects inter-procedural analysis. Let's see a more strict case shown in Figure~\ref{fig:fc-control}, where both sides of the branch return a concrete value. Although the branch condition in Line 2 clearly indicates that $x=2$, data-flow sensitive analysis cannot determine this. In such cases, control-flow sensitive analysis is necessary. By combining control-flow sensitive analysis with the branch condition, we can encode it into the return value, \ie $((x==2)\And4) \oplus ((x!=2)\And1)$.

Compared to symbolic execution, concolic execution's lack of global information presents a significant challenge, making interprocedural analysis considerably more difficult. This issue is particularly critical in the context of directed fuzzing, which contrasts with coverage-guided fuzzing.
Directed fuzzing requires a higher level of precision in path exploration, and any missteps can lead to lost or stuck during exploration. The challenge is to balance precision with concolic execution's limitations.

\begin{lstlisting}[caption=A loop example.,label=fig:loop-code, frame=shadowbox]
while(true) {
    char input;
    switch(input) {
        case 'X':
            break;
        case 'Y':
            assert(0);/*error*/
            break;
        case 'Z':
            break;
        default:
    }
}
\end{lstlisting}

\begin{figure}[t]
    \centering
    \includegraphics[width=0.6\columnwidth]{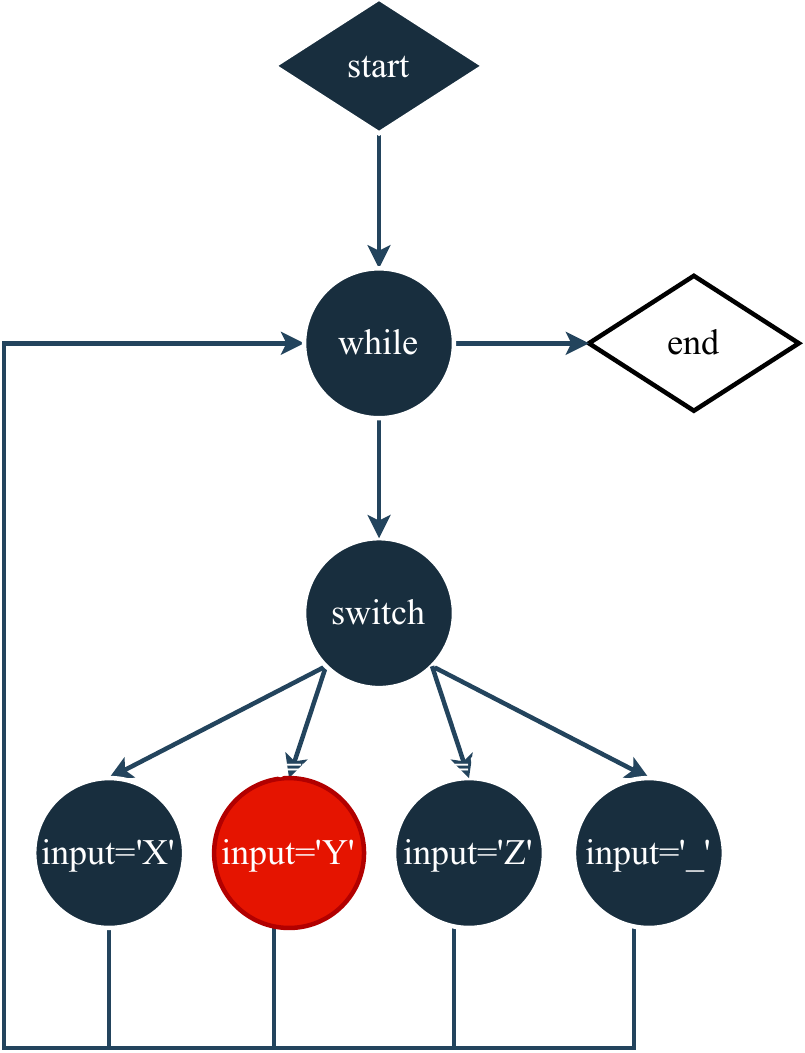}
    \caption{Control-flow graph of the loop example.}
    \label{fig:loop}
\end{figure}

\indent \textbf{Loop Unrolling.}
Loop and recursion are frequently used in programs, but traditional symbolic execution struggles to handle them effectively. The state explosion problem arises when symbolic execution forks a new path at every branch point (including the loop branch). This means that without a limiting measure, the number of paths grows infinitely at an exponential rate. Concolic execution's stateless nature avoids state explosion, but the problem of loops persists.

Consider the example shown in Figure~\ref{fig:loop}, where line 7 is the fuzzing target. A naive way to mark the relevant path would be to iteratively mark the basic blocks preceding the target basic block. If the target basic block is the red one, then all basic blocks except the end node would be marked as relevant in the search space. As a result, when we execute the switch statement, we lose direction to the final target node, because all successors of the switch statement are marked as relevant in the circular execution flow. Therefore, we need to design a special method for handling loop and recursion, as the naive method performs well for other statements except loop statement.
\begin{figure*}
    \centering
    \includegraphics[width=2\columnwidth]{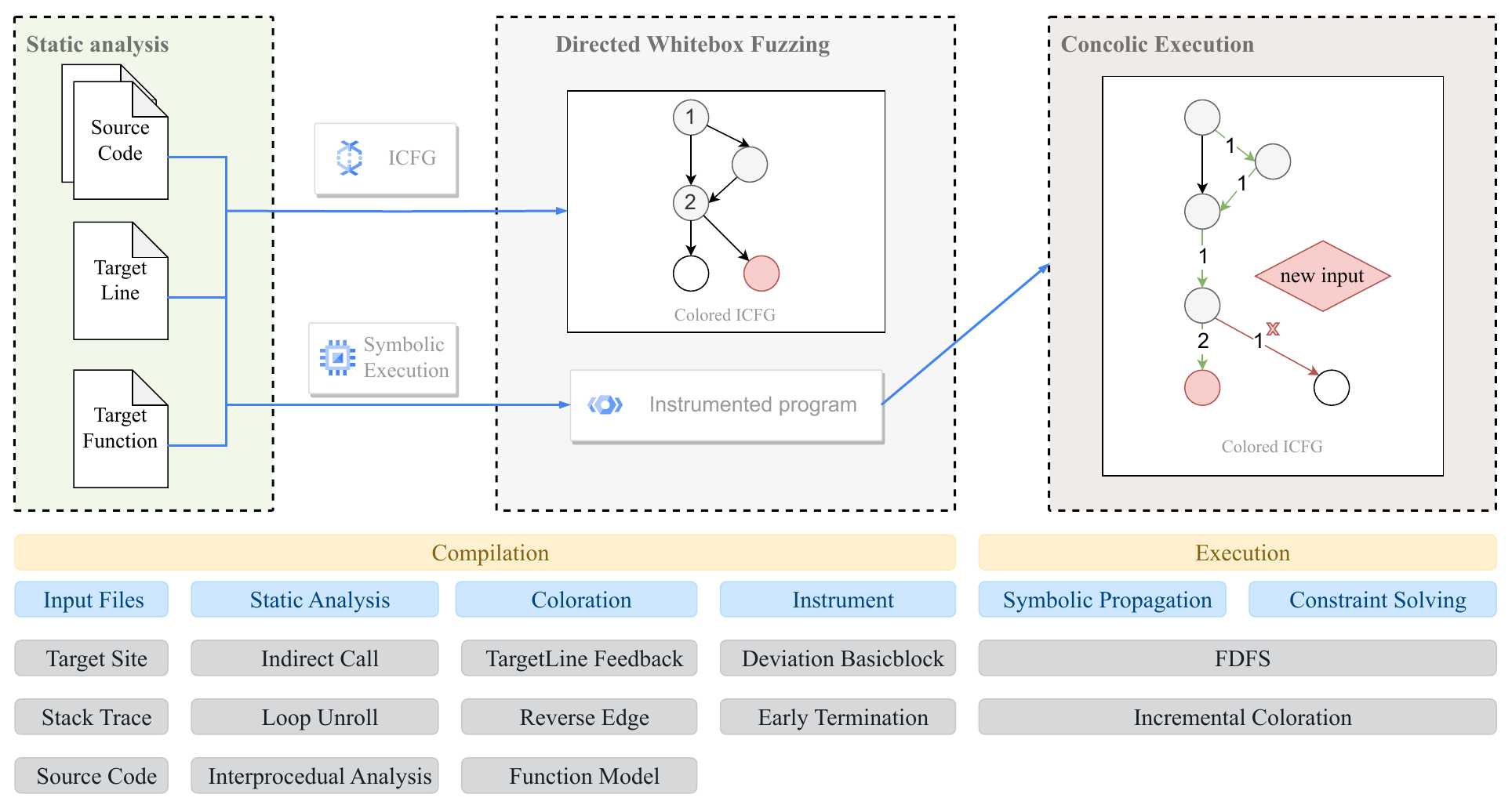}
    \caption{Architecture of ColorGo.}\label{fig:arch}
    \vspace{-0.1in}
\end{figure*}

\section{ColorGo}\label{sec:design}
In this section, we present the design of ColorGo, a directed whitebox fuzzer employing concolic execution and exploration space coloration for efficient crash reproduction. It addresses the low scalability of directed symbolic execution and the low precision of directed graybox fuzzing by introducing concolic execution into the directed fuzzing domain and overcoming the lack of global information and the inherent limitation of symbolic execution. As a result, it achieves high scalability. 

\subsection{Overview}
The overall architecture of ColorGo is depicted in Figure \ref{fig:arch}. In brief, we spread works in both compilation time and run time. At compilation time, we process source code with files that contain target lines and stack trace information, generating colored iCFG and instrumented programs. At runtime, we iteratively execute the generated instrumented program on the input set, at the same time generating new inputs added to the input set, until reach the target site. We now describe some important components of our approach. The detailed implementation will be described in the next section.

\subsection{Incremental Coloration} \label{sec:increment}
ColorGo is a tool that performs incremental coloration to restrict search scope and improve search performance. It uses global information to achieve this. The aim is to identify relevant code that needs to be explored, to avoid wasting time on paths that won't help to reach the target sites. 

To accomplish this, we convert directed fuzzing into a one-source, multi-targets graph search problem on iCFG. This involves marking the search scope using a process called coloration. The marking process is divided into two phases: static coloration and dynamic coloration. 

Static coloration is performed at compilation time. Given target lines(\eg main.c:5) and a function call chain extracted from the stack trace,  we do static coloration in an LLVM pass, where we can exploit the knowledge available in the compiler as part of global information. For each function, we color the target basic blocks which include the target line's corresponding instructions using debug information, or call instruction whose operand is the target function; Then, we iteratively color the predecessor basic blocks of target basic blocks. Finally, we get a connected subgraph of origin iCFG, which is called colored iCFG. 
The step is straightforward except we should specifically handle the indirect call and loop statement mentioned in Section \ref{sec:problem}. We integrate a conservative points-to-analysis tool~\cite{andersen1994program} to translate indirect function call into corresponding target line and feedback to the beginning of our framework. Besides, we stop the iterative coloration when detecting the reverse edge, \ie edges from the back basic block to the front basic block.  

Dynamic coloration is performed at runtime and uses path constraint feasibility information as a supplement to global information. While graybox fuzzer~\cite{huang2022beacon} designs carefully to balance the precision and efficiency when pruning infeasible paths. The concolic execution natively supports runtime infeasible path pruning. At each key branch point, the instrumented program extracts the path constraint that points to the colored side and sends it to the backend solver to derive a solution. If the path constraint is infeasible, its subtree in colored iCFG is sliced away and won't be explored again, which is called incremental coloration.

The coloration information will be later used in directed concolic execution to restrict search space and achieve a high-performance search strategy, we discuss this in the next section.

\subsection{Compilation-based Concolic Execution}
The key component of ColorGo is the directed concolic executor, which is implemented as an instrumented program.

In this work, we choose compilation-based concolic execution for these reasons: Firstly, compared with the runtime instrumentation, the injected code can run seamlessly with the application code, eliminating the need for switching between the target and an interpreter or an attached observer, achieving low run-time overhead. Secondly, we require the assistance of code analysis information and high-level knowledge generated by the compiler to generate global information to guide the coloration. Thirdly, compared to source-to-source translation, the compiler intermediate representation (IR) level instrumentation simplifies the integration of concolic execution capability as we only need to handle a limited instruction set.

We instrument the direct logic into the program when doing compilation. The most important question is how our directed logic works. For the search strategy, we borrow the concept deviation basic block from Windranger~\cite{du2022windranger}, where the execution trace starts to deviate from the target sites. For example, in the middle Colored iCFG in Figure~\ref{fig:arch}, node 1 is not a deviation basic block, for its both siblings are colored nodes. But node 2 is a deviation basic block for one of its siblings is not a colored node. We only generate constraint solving when current execution trace starts to deviate from target sites. Take the right Colored iCFG in Figure~\ref{fig:arch} as an example, to reach the target site we conduct two times of execution, which is marked on the arrows as number 1, 2. In every deviation basic block we generate a new input on which the program will reach the colored basic block. After one time of "corrections", the program executes on $new input$ and will finally reach target red node. 
We call this search strategy Fast Depth First Search (FDFS), to reach the target node with minimal time for input generation and program execution. Suppose we do not use this method, a naive DFS generate input towards every colored side, which results in generating an unnecessary solution at the first branch point. This time wasting is more severe in real-world programs. 
For search scope restriction, we terminate the execution when runs to the end node of iCFG, \ie both siblings are uncolored, or at branch points with an unfeasible path condition. 

\subsection{Interprocedural Analysis}
Unlike coverage-guided fuzzing which negates the branch conditions along the way, directed fuzzing needs to execute a more complex logic to selectively explore path. As discussed in Section~\ref{sec:problem}, we need to implement both data-flow sensitive and control-flow sensitive methods. The data-flow sensitive analysis follows SYMCC. When do function calls, we propagate the symbolic function parameter throughout the subprocess instructions to construct the corresponding symbolic expression in the called function. Finally, we register the return expression for later use. However, the control-flow sensitive feature is harder to realize in the concolic executor. At each branch point, the concolic executor only sends the current branch condition to the symbolic backend, and the complete path constraint is collected during runtime. Therefore, if we want to encode path constraints to return value during compilation time, we need to conduct complicated static analysis. 
This is similar to the function summary technique in symbolic execution. The function summary can be defined as a disjunction of formulas, which is present in the form that combines a conjunction of constraints on the inputs and a conjunction of constraints on the outputs.
For example, the summary for the function $f$ in Figure~\ref{fig:fc-control} could be $(x = 2 \wedge ret = 4) \vee (x \neq 2 \wedge ret = 1)$. 
SMART~\cite{godefroid_compositional_nodate}uses static analysis to generate function summaries to assist the inter-procedural analysis of symbolic execution, which effectively alleviates the path explosion problem caused by too many function calls. However, This heavy global static analysis and status management obey the design concept of concolic execution, and result in low scalability. 

To balance the scalability and effectiveness, we propose a partial function model. While modeling all user-defined functions in the program introduces great complexity to our design, we only model functions in C standard library, which is considered simple but important. To normally reason most of the program, we build a summary for some important functions in C standard library (\eg strlen and strchr). We will discuss this in detail in Section~\ref{sec:implementation}.

\section{Implementation}\label{sec:implementation}
We discuss the detailed implementation of ColorGo in this section. We built ColorGo using LLVM pass, written from SYMCC. The pass consists of roughly 500 edited lines of C++ code. In this pass, we process LLVM intermediate representations one by one at both the module and function levels. To build the instrumented program, the instrumentation performs only once along with the compilation. To conduct directed concolic execution, the instrumented code will be executed over and over again until reach target site or all colored path are explored. 
The instrumented code reserves the behavior of the original program but invokes constraint solver when deviating from target execution path to generate new input that corrects the execution. We propagate the symbolic expression along the instructions by adding around each LLVM IR with calls to symbolic handling implemented in the run-time support library as SYMCC does. 
Our directed characteristic is mainly reflected on the time we initiate constraint solving. We build a concolic executor as a scheduler that picks the next input to execute on, written in a shell script, and comprises another 200 lines. The low code volume shows the flexibility of our system, thus it's easy to deploy a new directed logic based on our implementation with little work.

\subsection{Instrument at compilation time}
To derive instrumented binary, we conduct instrumentation at compilation time. The input of our approach source code files, target lines, and functions in the stack trace. Note that the target lines in a feedback version by points-to-analysis~\cite{andersen1994program}. In general, We will compile the source code of the program under test to an instrumented binary.

First, we read the input and register the target lines and functions for each function in a map. Second, for each function, we process each instruction in order, if we identify that the instruction maps to target line according to debug info or the instruction is a call/invoke instruction and its operands is target function, the basic block to which the instruction belongs would be marked as target basic block. Then we do a back propagation in control-flow graph which iteratively mark the predecessors of target basic block. Finally, we will mark the root node in CFG. After the coloration of all functions, we get a connected subgraph of origin iCFG which is called Colored iCFG. 
We specially handle loop statements to avoid coloring all basic blocks in the loop body, to be specific, we stop the propagation when detecting inverse edge (from back basic block to front basic block), which can effectively solve the loop pollution problem we introduce in Section~\ref{sec:problem}.

Strictly speaking, we will process all instructions twice at compilation time, at the first time we get colored iCFG and the second time we insert calls to symbolic backend to generate new input according to the colored iCFG. The execution trace diverges at every branch point, thus, we should insert check logic around every conditional branch/switch statement. If a function has an empty target basic block set, we will skip the check for there is no distinction among all basic blocks, which is a common case in subprocess.
\begin{itemize}
\item For conditional branch statements, the check logic acts belike: If both sides point to a non-colored basic block, execution terminates early. If both sides point to the colored basic block, skip check. If one of the sides points to a non-colored basic block and the another one points to a colored basic block, we send the symbolic expression of branch constraint, concrete value of branch constraint, and a boolean value which represent the wanted value of branch constraint to the symbolic backend, and symbolic backend will check the equality of concrete value and wanted value, if equal, add this branch constraint according to the concrete value to path constraint, if not, initiate a constraint solving to satisfy the equation of path condition and the wanted value.
\item For switch statement, the check logic acts like the conditional branch statement in general, but there is a little difference: when constructing the constraint, we can not directly extract the case constraint from the operand, especially for the default case, which needs to be handled manually. If at least one of the sides points to a non-colored basic block, we will produce solutions for each (not one of) case which points to a colored basic block.
\end{itemize}

\subsection{The Placement of the instrumentation in the Compiler Pipeline}
The placement of the instrumentation pass is nontrivial. We place the pass in early the pipeline to achieve maximum structure similarity to the source code, in order to best map to the input target information. For instance, the compiler may merge some functions when doing optimization, after which we can not extract the origin function information and it is hard to map the optimized version to the original execution flow. So directed fuzzer is lost here if we instrument the optimized version. 

\subsection{Explore at runtime}
After the compilation, we derived a symbolized binary that drives the program to the target site. At runtime, we just repeatedly execute the binary and collect generated input. The input helps to correct current execution traces to a specified path leading to target sites. Recall the Section~\ref{sec:increment}, we cast directed fuzzing into one source, multi-targets graph search problem on iCFG. At compilation time, we define the search scope by coloration statically and embed deviation-correction and early-termination logic into the program at compilation time. Now the final question comes: How do we schedule the newly generated input to realize our Fast Depth First Search (FDFS) strategy?

The answer is straightforward, we maintain a stack as the input pool. Every newly generated input is pushed to the top of the stack and tends to be executed immediately. The FDFS acts like a persistent detective who wants to explore deep into the program as fast as possible. It terminates as soon as know there is no way to target sites, \ie at branch point whose siblings are all uncolored, or when current path constraint is infeasible. It only corrects the execution when necessary (in deviation basic block) because correction means re-executing the program.
Besides, we avoid generating repeated input by a map, which is practical when at branch point more than one side are colored and generating inputs run to each other over and over again when executing.

\subsection{Compilation Boundary and Function Model}
The compilation-based approach also has its intrinsic drawback, that is, not all programs can be recompiled. For example, some system libraries and third-party libraries are not typically recompiled by application developers. The symbolic execution would degrade to concrete execution when running uninstrumented code. Latest work~\cite{coppa2022symfusion} proposes a hybrid instrumentation approach for concolic execution which instruments internal code at compilation time and external code at runtime. But SYMFUSION incurs an average of 3X slowdown. In fact, when doing fuzzing, we only focus on program under test, and the most of symbolic state loss because of uninstrumented code can be relieved by function model, which is also preserved by SYMFUSION. We model important functions in C standard library and additionally conduct our data-flow and control-flow sensitive analysis in function models to produce function summary (\ie encode the path condition into the return expression). For example, to model function $const char *strchr(const char *str, int c)$, which find char $c$ in string $str$ and return the position of $str$ where $c$ first happens, we set the return expression as $((s[0]==c)\And s[0]) \oplus ((s[i]==c)\And s[i])$.

\section{Evaluation}\label{sec:evaluation}
In this section, we evaluate ColorGo using real-world programs and answer the following questions:
\begin{itemize}[leftmargin=*]

\item  \textbf{RQ1:} How fast can ColorGo reach target sites?
\item  \textbf{RQ2:} How fast can ColorGo expose vulnerabilities?
\item  \textbf{RQ3:} How does every component in ColorGo contribute to the overall performance?
\item  \textbf{RQ4:} What's the runtime overhead introduced by ColorGo's instrumentation?

\end{itemize}

\begin{table}[]
    \vspace{0.15in}
    \footnotesize
    \centering
    \caption{\small Real-world benchmark programs used in the evaluation.}
    \label{tab:bm-info}
    \begin{tabular}{l|l|r|r}
        \toprule[2pt]
        \textbf{Project} & \textbf{Program} & \textbf{\# Version}    & \textbf{\# Input Format}        \\
        \midrule[1pt]
        \textit{jasper}               & imginfo              &  2.0.12   & Image \\
        \textit{lame}      & lame                  & 3.99.5        & Audio  \\
        \textit{binutils} & cxxfilt & 2.26        & Txt \\
        \bottomrule[2pt]
    \end{tabular}
\end{table}

\subsection{Evaluation Setup}
\noindent\textbf{\textit{BaseLine.}}
We compare ColorGo with a state-of-the-art directed graybox fuzzer AFLGo. Which is publicly available by the time of writing this paper.

\noindent\textbf{\textit{Evaluation Criteria.}}
We use two criteria to evaluate the performance of different fuzzing techniques.
\begin{itemize}[leftmargin=*]
\item Time-to-Reach (TTR) is used to measure the time fuzzer used to generate the first input on which the program can reach the target site.
\item Time-to-Expose (TTE) is used to measure the time fuzzer used to generate the first input on which the program can reproduce known venerability.
\end{itemize}

\noindent\textbf{\textit{Evaluation Datasets.}}
We use real-world programs from the following datasets as the evaluation benchmarks:
\begin{itemize}[leftmargin=*]
\item UniBench~\cite{li2021unifuzz} is a recent dataset proposed for evaluating fuzzing techniques. It consists of 20 real-world programs from 6 different categories, categorized based on the input file type. From this dataset, we selected some programs to measure the Time-to-Reach (TTR) of baseline and our work. These benchmarks are used to address \textbf{RQ1}.
\item AFLGo Test Suite~\cite{AFLGOtestsuite} is a collection of programs with n-day vulnerabilities, which was used in the experiments of AFLGo~\cite{AFLGo}. This test suite has been utilized in multiple research studies~\cite{chen2018hawkeye,du2022windranger} to evaluate DGF techniques. These benchmarks are used to address \textbf{RQ2-4}.
\end{itemize}

\noindent\textbf{\textit{Experiment Settings.}}
We conducted our evaluations on the machine equipped with Intel Xeon Gold 5218R CPU with 20 cores, using Ubuntu 20.04.6 LTS as the operating system. During experiments, each fuzzer instance runs in a docker container~\cite{Docker} and binds to one CPU core.

The baseline DGF was repeated 10 times with a time budget of 24 hours. Our method does not contains randomness, and therefore statistical evaluation is unnecessary. We only evaluate our method once. The detail of the real-world benchmark programs we used is shown in Table~\ref{tab:bm-info}.

\begin{table*}[]
    \small
    \centering
    \caption{\small Time-to-Reach results on programs from UniBench.}\label{tab:ttt}
    \begin{tabular}{clllllllll}
        \toprule[1pt]
        \multirow{2}{*}{\textbf{Prog.}} & \multicolumn{1}{c}{\multirow{2}{*}{\textbf{Targets}}} & \multicolumn{2}{c}{AFLGo} & & \multicolumn{5}{c}{ColorGo}          \\
                                 & \multicolumn{1}{c}{}                         & Runs       & $\mu$TTR  &   & Executions & TTR    & TET      &Solves & $\mu$ST    \\
        \midrule[0.5pt]
        \multirow{3}{*}{imginfo} & jpc\_cs.c:316                                & ——         & T.O       &   & 2    & 275 ms      & 160 ms     & 1      & 406 $\mu$s    \\
                                 & bmp\_dec.c:373                               & ——         & T.O       &   & 16   & 314169 ms   & 312656 ms  & 1      & 1098 $\mu$s   \\
                                 & jp2\_cod.c:275                               & ——         & T.O       &   & 2    & 266 ms      & 90 ms      & 2      & 691 $\mu$s    \\
        \midrule[0.5pt]
        \multirow{3}{*}{lame}    & get\_audio.c:1280                            & 10         & 26.7s     &   & 2    & 587 ms      & 190 ms     & 1      & 897 $\mu$s    \\
                                 & get\_audio.c:1285                            & 10         & 51.3s     &   & 2    & 594 ms      & 198 ms     & 1      & 1191 $\mu$s   \\
                                 & get\_audio.c:692                             & 10         & 64.05s    &   & 2    & 1228 ms     & 784 ms     & 1      & 195233 $\mu$s \\
        \bottomrule[1pt]
    \end{tabular}
    \vspace{-0.1in}
\end{table*}

\begin{table*}[]
  \small
    \centering
    \caption{\small Time-to-Expose results on AFLGo test suite.}\label{tab:ttr}
    \begin{tabular}{clllllllllll}
        \toprule[1pt]
        \multirow{2}{*}{\textbf{Prog.}}        & \multicolumn{1}{c}{\multirow{2}{*}{\textbf{Targets}}} & \multicolumn{2}{c}{AFLGo} & &\multicolumn{5}{c}{ColorGo}                        \\
                                        & \multicolumn{1}{c}{}                          & Runs       &$\mu$TTE   &   & Executions & TTE   & TET    & Solves & $\mu$ST & ETE \\
        \midrule[0.5pt]
        \multirow{4}{*}{binutils}       & 2016-4487(4489)                               & 10         & 40.6s     &   & 5        & 308 ms   & 180 ms   & 4      & 658 $\mu$s  & 5             \\
                                        & 2016-4488                                     & 10         & 1s        &   & 5        & 927 ms   & 710 ms   & 9      & 746 $\mu$s  & 5             \\
                                        & 2016-4490                                     & 10         & 13s       &   & 3        & 122 ms   & 60 ms    & 5      & 494 $\mu$s  & 3           \\
                                        & 2016-4492                                     & 10         & 44.8s     &   & 4        & 444 ms   & 172 ms   & 4      & 399 $\mu$s  & 4             \\
        \bottomrule[1pt]
    \end{tabular}
    \vspace{-0.1in}
\end{table*}

\subsection{Performance on Reaching Target Site} \label{sec:rq2}
We conduct the evaluation on two kinds of open-source real-world programs from UniBench. Table~\ref{tab:ttt} shows the results.
\begin{itemize}[leftmargin=*]
\item Jasper is a collection of software (\ie a library and application programs) for the coding and manipulation of images. This software can handle image data in a variety of formats. One such format supported by Jasper is the JPEG-2000 format defined in ISO/IEC 15444-1. Our target covers three formats of image, including \textit{jpc}, \textit{bmp}, and \textit{jp2}.
\item LAME is an MP3 encoding tool. The goal of the LAME project is to use the open-source model to improve the psycho acoustics, noise shaping, and speed of MP3. 
\end{itemize}

The metric we use to measure the performance is the time cost to reach the selected target sites. Additionally, we provide the mean solve time which represents the solver used to produce a solution. On all target sites, ColorGo outperforms benchmark fuzzer and achieves the shortest $\mu$TTR. Overall, even discard data that AFLGo timeout (>24h). In terms of mean TTR, ColorGo outperforms DGF (AFLGo) by \textbf{50×}. The result shows that our concolic execution, which takes both precision and efficiency into account, has better capability to reach the target site than DGF.

We use the same initial inputs in UniBench for all experiments, the number 16 of runs in this table shows the demand of sorting the inputs in the initial queue, while our work tries them in order.

We add a new metric called Total Execution Time (TET) which separately counts the time used on program execution, the subtraction of TTT and TET represents time for concolic executor to build up, including some file I/O operations and the maintenance of input queue. We can see more than half of the time spent on the starting process in our experiment, but when the total time is longer, the proportion of starting time is smaller.

\subsection{Performance on Exposing specific vulnerability}
In this section, we evaluate the vulnerability reproduction performance by the time used to trigger specific crushes. Vulnerability reproduction is more in line with the actual application scenario, and the vulnerability report always includes stack trace of the error, which records the function call stack. The function call chain further restricted the search scope at the level of Call Graph. We search the information of CVEs we reproduce on the official website and record the function call stack for our later use. Table~\ref{tab:ttr} shows the results.

The program we selected to evaluate our work is:
\begin{itemize}[leftmargin=*]
\item Binutil. The GNU Binutils are a collection of binary tools. We evaluate c++filt, a filter to demangle encoded C++ symbols.
\end{itemize}

We add a new metric called Early Termination Executions (ETE) to illustrate the effectiveness of our coloration. The percentage of early termination runs in all runs is \textbf{100\%}, which proves our coloration does help to avoid wasting time on irrelevant code exploration. 

Except for CVE-2016-4488, which is considered very easy to expose ($\mu$TTE < 1s), ColorGo significantly outperforms other tools by \textbf{100×} faster to expose them. The result shows that ColorGo performs best when the path is highly specified. In such cases, we can maximize the effectiveness of precise seed generation.

\begin{table}[h]
  \small
    \centering
    \caption{\small Mean Execution Time on AFLGo test suite. ET = early termination}\label{tab:differential}
    \begin{tabular}{clllllllll}
        \toprule[1pt]
        \multirow{2}{*}{\textbf{Prog.}}        & \multicolumn{1}{c}{\multirow{2}{*}{\textbf{Targets}}} & \multicolumn{3}{c}{Mean Execution Time}         \\
                                        & \multicolumn{1}{c}{}                         & \textbf{Default} & \textbf{No ET} & \textbf{Pure execution} \\
        \midrule[0.5pt]
        \multirow{4}{*}{binutils} & 2016-4487(4489)                                & 36 ms   & 40ms                 &  16ms              \\
                                        & 2016-4488                                     & 142 ms  & ——                   &   47ms             \\
                                        & 2016-4490                                     & 20 ms   & 38ms                 &   17ms             \\
                                        & 2016-4492                                     & 43 ms   & 48ms                 &   16ms            \\
        \bottomrule[1pt]
    \end{tabular}
    \vspace{-0.1in}
\end{table}

\subsection{Impact of Different Components}
To investigate the impact of different components in ColorGo, we disable each component individually and conduct experiments on the same targets selected from AFLGo Test Suite as in Section~\ref{sec:rq2}.

\subsubsection{Impact of Search Scope Restriction}

We conducted an experiment to study the impact of search scope restriction. To do this, we disabled the early termination mechanism, which allowed the program to execute outside the colored space. The out-of-colored execution is useless and time-wasting and may produce new input which results in irrelevant execution. The disabled one follows almost the same execution path as the original one. We compare the average execution time to show the speedup for each execution of our method.
Table~\ref{tab:differential} shows the results. We can observe that disabling the early termination causes an increase (>10\%) on TTE, which means the search scope restriction has a significant impact on the performance of ColorGo.
The increased degree of TTE depends on the depth of our target. If the target is located near the beginning of the program, the effect of early termination will be more significant. For CVE-2016-4490, by conducting coloration and path pruning, we reduced the average execution time by 50\%.
Besides, the CVE-2016-4488 failed to be reproduced because of the huge change in execution path caused by disabling part of the coloration, which also demonstrates our coloration can help the path-finding of directed fuzzer.

\subsubsection{Impact of Search Strategy}
To validate the effectiveness of our search strategy, we compare our FDFS with the implementation of naive DFS which disables the concept of deviation basic block, which issues constraint solving towards every colored side along the execution path. The result is presented in Table~\ref{tab:searchstrategy}, it is clear that the number of useless solutions has increased a lot (100×) as we collect all potential input along the way in coloration scope. The increase of useless solutions and inputs results in the growth of the number of executions, which significantly degrade the performance. The degradation of performance proves our FDFS outperforms naive DFS a lot.  

\begin{table*}[]
  \small
    \centering
    \caption{\small Performance of different search strategies on AFLGo test suite.}\label{tab:searchstrategy}
    \begin{tabular}{clllllllllll}
    \toprule[1pt]
    \multirow{2}{*}{\textbf{Proj.}}    & \multicolumn{1}{c}{\multirow{2}{*}{\textbf{Targets}}} & \multicolumn{5}{c}{FDFS}                              & \multicolumn{5}{c}{DFS}                               \\
                              & \multicolumn{1}{c}{}                         & Executions & \makecell{Mean \\Execution Time} & Solves & $\mu$ST     & TTE    & Executions & \makecell{Mean \\Execution Time} & Solves & $\mu$ST     & TTE    \\
    \midrule[0.5pt]
    \multirow{4}{*}{binutils} & 2016-4487(4489)                              & 5    & 36 ms               & 4      & 658 $\mu$s & 308 ms & 39    & 47 ms               & 365      & 462 $\mu$s & 25450 ms \\
                              & 2016-4488                                    & 5    & 142 ms              & 9      & 746 $\mu$s & 927 ms & 25    & 43 ms              & 112      & 559 $\mu$s & 5810 ms \\
                              & 2016-4490                                    & 3    & 20 ms               & 5      & 494 $\mu$s & 122 ms & 8    & 41 ms               & 39      & 513 $\mu$s & 1009 ms \\
                              & 2016-4492                                    & 4    & 43 ms               & 4      & 399 $\mu$s & 444 ms & ——   & ——                  & ——     & ——     & ——     \\
    \bottomrule[1pt]
    \end{tabular}
\end{table*}

\subsection{Instrumentation Overhead}
Concolic execution obtains symbolic capability by instrumentation, which causes additional runtime overhead. To evaluate the runtime overhead, we run the same input against three versions of the benchmark program in Section~\ref{sec:rq2}. One is the vanilla version without any instrumentation, and the other is instrumented by ColorGo, where we add symbolic expression propagation and constraint solving but disable the early termination to ensure that we run the same paths on two versions. Additionally, we also present the default data we obtain in Section \ref{sec:rq2} by experimenting with the native implementation of ColorGo as a reference. The results are shown in Table \ref{tab:differential}, we compare the default mean execution time, mean execution time disabled early termination, and mean execution time of pure execution. We have observed that ColorGo can cause up to a 67\% runtime overhead which reduces to 62\% on average. If early termination is used, this percentage could be decreased to 50\%. The cost of instrumentation is negligible compared to the cost of interpretation.

\section{Related Work}
There are two threads of work in the literature related to ColorGo, \ie directed graybox fuzzing and hybrid fuzzing.
In this section, we introduce these works accordingly.

\subsection{Directed Graybox Fuzzing}
Nowadays, many new works emerge to optimize directed graybox fuzzing, including fitness metrics~\cite{ye2020rdfuzz,zhu2021regression,chen2018hawkeye,wang2020typestate} to conduct seed prioritization, make fuzzing optimization including input optimization~\cite{jain2018tiff,you2017semfuzz,zong2020fuzzguard,huang2022beacon}, power scheduling~\cite{chen2018hawkeye, zhu2021regression,liang2019sequence,liang2020sequence}, mutator scheduling~\cite{chen2018hawkeye, li2020v, you2017semfuzz, situ2019energy}, and mutation operations~\cite{ye2020rdfuzz, wang2020typestate,you2017semfuzz, jain2018tiff}, to make directed fuzzing more directed.

Regarding search scope restriction, we prove that there is an intersection between whitebox directed fuzzing and graybox directed fuzzing, and there may be other opportunities to learn from each other for eventual performance gains.
To enhance our search strategy design and make more informed decisions about which constraints to solve and which inputs to execute, we can assign a weight to each edge in the iCFG. This takes into account various influencing factors. This weighting system is akin to the fitness metrics used in Directed Greybox Fuzzing (DGF). DGF employs a fitness metric to gauge how closely the current fuzzing aligns with the fitness goal.
While early iterations of DGF only considered distance on the iCFG, numerous variants have been developed to refine the fitness metric. For instance, TOFU~\cite{wang2020tofu} defines its distance metric as the number of correct branching decisions required to reach the target. RDFuzz~\cite{ye2020rdfuzz}, on the other hand, combines distance with the execution frequency of basic blocks to prioritize seeds.
AFLChurn~\cite{zhu2021regression} assigns a numerical weight to a basic block based on its recent changes or frequency of alterations. WindRanger~\cite{du2022windranger}, meanwhile, factors in deviation basic blocks. 

The similarity metric is proposed by Hawkeye~\cite{chen2018hawkeye}, which is the degree of overlap between the current status and target status from a certain aspect, including execution trace similarity~\cite{chen2018hawkeye}, statement sequence similarity~\cite{liang2019sequence,liang2020sequence}, and so on. Similarity metric is also used for specific bug detection such as use-after-free bug and other memory-related bugs~\cite{chen2020savior,nguyen2020binary}.

Deep learning has also played a role in vulnerable probability prediction at the level of function and basic  block~\cite{li2020v,zhao2020suzzer,joffe2019directing}. The probability-based metric allows for the combination of seed prioritization and target identification. This enables directing fuzzing towards potentially vulnerable locations, without being dependent on the source code.

These adaptations demonstrate the evolution and sophistication of the fitness metric, enabling more nuanced and effective fuzzing strategies.

\subsection{Hybrid Fuzzing}
Hybrid fuzzing~\cite{noller2020hydiff,chen2020savior,noller2018badger,kim2019poster,liang2020sequence} combines symbolic execution and graybox fuzzing, to utilize the advantages of each technique. In this scenario, symbolic execution acts as an assistant by solving condition of branch that is hard to cover by graybox fuzzing. Hybrid fuzzing also aims to combine the precision of DSE and the scalability of DGF to mitigate their individual weaknesses by selectively using symbolic execution, under the observation that DGF tends to explore branches with simpler path constraints while DSE is geared towards solving complicated path constraints. However, it does not solve the problems of both techniques fundamentally as our work has done. Two components intertwining in the system also introduce complexity and make the system more cumbersome and inefficient.

\section{Conclusion}
In this paper, we propose a directed whitebox fuzzer that utilizes compilation-based concolic execution. Our innovative approach aims to address the inefficiency and imprecision issues found in existing graybox and whitebox directed fuzzers.
We introduce ColorGo, a tool designed to effectively reach target sites by iteratively concretely executing an instrumented program. To enhance the directed whitebox fuzzing, we propose incremental coloration to restrict the exploration scope and overcome the limitations of concolic execution.
Furthermore, we conducted a comparative experiment with state-of-the-art directed graybox fuzzing. Our results demonstrate that our method outperforms it by reaching the target sites 50 times faster and exposing vulnerabilities 100 times faster. Additionally, we thoroughly investigate the effect of different parts of our design on the final performance.


\section{Data Availability}
The source code and detailed experiment are available on the website~\cite{datapublic}. Besides, we demonstrate the guidelines for the reproduction of all experiment results.

\bibliographystyle{plain}
\bibliography{main}
 
\end{document}